\pdfoutput=1
\documentclass[superscriptaddress,twocolumn,floatfix]{revtex4}
\usepackage{graphicx}
\usepackage{amsmath}
\usepackage{color}

\begin{document}

\title{Coherent control of optical activity and optical anisotropy of thin metamaterials}

\author{Seyedmohammad A. Mousavi}
\affiliation{Optoelectronics Research Centre and Centre for Photonic Metamaterials, University of Southampton, SO17 1BJ, UK}

\author{Eric Plum}
\email{erp@orc.soton.ac.uk}
\affiliation{Optoelectronics Research Centre and Centre for Photonic Metamaterials, University of Southampton, SO17 1BJ, UK}

\author{Jinhui Shi}
\affiliation{Optoelectronics Research Centre and Centre for Photonic Metamaterials, University of Southampton, SO17 1BJ, UK}
\affiliation{Key Laboratory of In-Fiber Integrated Optics of Ministry of Education, College of Science, Harbin Engineering University, Harbin 150001, China}

\author{Nikolay I. Zheludev}
\email{niz@orc.soton.ac.uk}
\homepage{www.nanophotonics.org.uk}

\affiliation{Optoelectronics Research Centre and Centre for Photonic Metamaterials, University of Southampton, SO17 1BJ, UK}
\affiliation{Centre for Disruptive Photonic Technologies, Nanyang Technological University, Singapore 637378, Singapore}

%\date{\today}

\begin{abstract}
The future fibre optic communications network will rely on photons as carriers of information, which may be stored in intensity, polarization or phase of light. However, processing of such optical information usually relies on electronics. Aiming to avoid the conversion between optical and electronic signals, modulation of light with light based on optical nonlinearity has become a major research field, but real integrated all-optical systems face thermal management and energy challenges. On the other hand, it has recently been demonstrated that the interaction of two coherent light beams on a thin, lossy, linear material can lead to large and ultrafast intensity modulation at arbitrarily low power resulting from coherent absorption \cite{LSA_2012_MMCoherentAbsorption}. Here we demonstrate that birefringence and optical activity (linear and circular birefringence and dichroism) of functional materials can be coherently controlled by placing a thin material slab into a standing wave formed by the signal and control waves. Efficient control of the polarization azimuth and ellipticity of the signal wave with the coherent control wave has been demonstrated in proof-of-principle experiments in different chiral and anisotropic microwave metamaterials.
%Furthermore, we demonstrate coherent intensity modulators and a novel spectroscopic technique allowing the selective excitation of electric and magnetic resonances. All phenomena are intensity independent, ultrafast and can be scaled across the electromagnetic spectrum promising 10s of THz bandwidth at optical frequencies.
\end{abstract}

\maketitle

%\section{Introductory paragraph}

%[TO DO: reference arXiv on coherent spectroscopy]
%[NOTE to NIZ: Fig. 5 is needed to make the paper complete. It allows us to speak about BOTH output beams.]

\begin{figure} [tb]
\includegraphics[width=80mm]{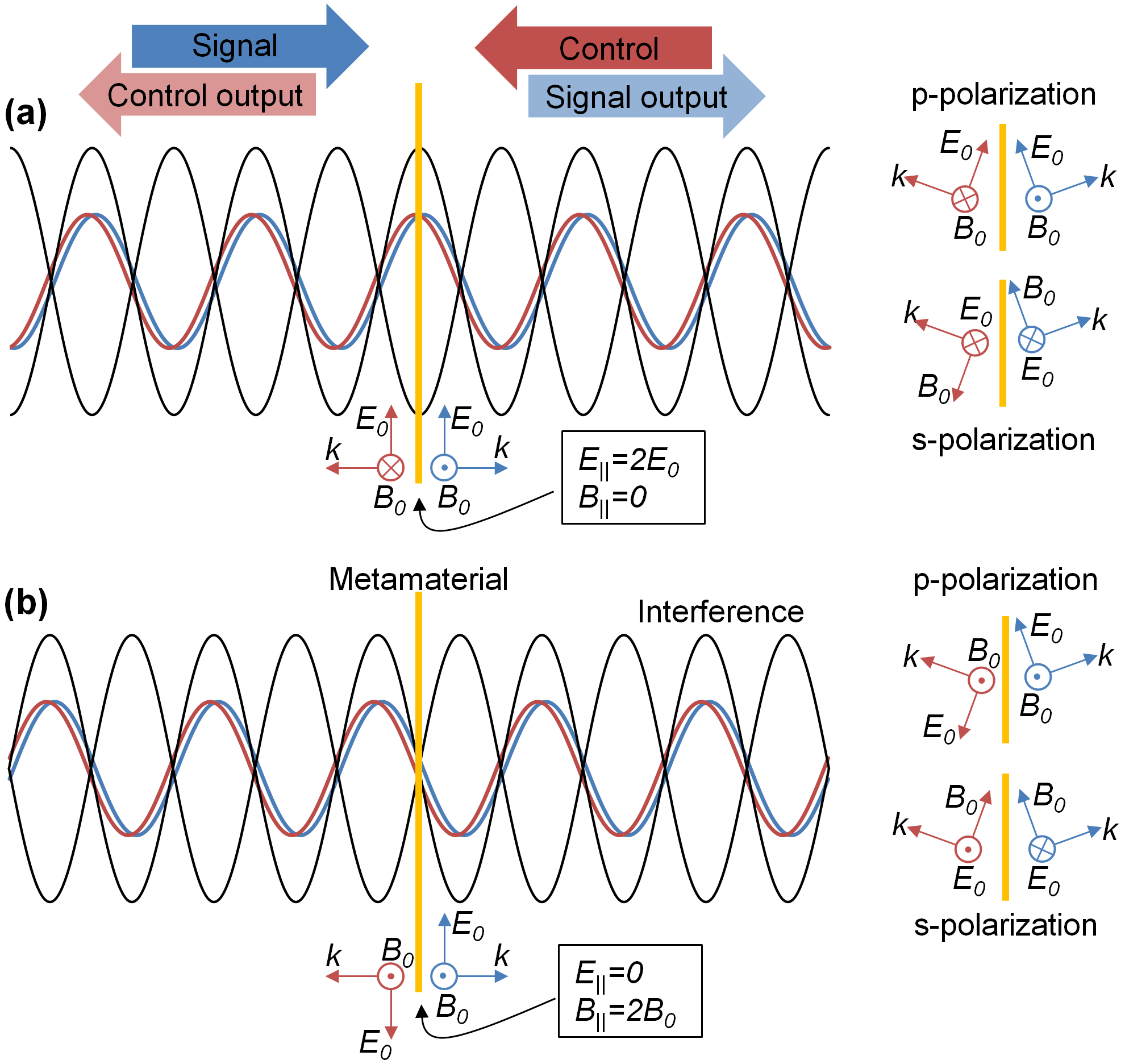}
\caption{\label{Fig_concept}
\textbf{Coherent control of metamaterial functionalities.} Two coherent counterpropagating beams ``Signal" and ``Control" form an interference pattern. In the limiting cases, a functional material of substantially subwavelength thickness can be placed at a position of (a) constructive interference or (b) destructive interference of the incident electric fields $E_0$ leading to enhanced or vanishing electric excitation of the material, respectively. Similarly, (a) destructive and (b) constructive interference of the incident magnetic fields $B_0$ leads to vanishing or enhanced magnetic excitation of the metamaterial, respectively. Insets illustrate the corresponding cases of constructive and destructive interference for p-polarization and s-polarization at oblique incidence.
}
\end{figure}

%\section{Introduction}
%Coherent interaction of electromagnetic waves has led to applications from phased array antennas to manipulation of light distribution and quantum states of matter [REFs 4-11 Coherent absorption paper].

Two coherent counterpropagating beams form a standing wave interference pattern consisting of nodes and anti-nodes of electric and magnetic field, see Fig.~\ref{Fig_concept}. While functional optical materials have historically been thick compared to the wavelength, the rise of metamaterials has generated a large number of functional materials of substantially subwavelength thickness across the electromagnetic spectrum from microwaves to optical frequencies. For example, ultra-thin absorbers \cite{LSA_2012_MMCoherentAbsorption}, linear polarizers \cite{WireGrid_visible}, wave plates \cite{OE_2009_THz_QuarterWavePlates}, optically active materials \cite{PRL_KuwataGonokami_2005_GiantOpticalActivity, APL_Plum_2007_GiantOpticalGyrotropy, OL_Decker_2007_CircDichroism, APL_Plum_2008_Extr3D, PRL_Plum_2009_Extr3D, PRB_Plum_2009_chiralNIM, PRB_2009_Zhou_TwistedCrosses_EffParameterRetrieval, OL_Decker_2009_TwistedCrosses}, phase gradient surfaces \cite{Science_2011_Capasso_PhaseGradients} and lenses \cite{LSA_2013_visible_phase_gradient_lens} have been demonstrated. This provides a new opportunity to control the associated optical phenomena coherently by placing the functional material at a node or anti-node of the standing wave interference pattern. As illustrated by Fig.~\ref{Fig_concept}, a material placed at an electric anti-node will experience twice the electric excitation field compared single beam illumination, while no electric excitation can occur if the material is placed at an electric node. Similarly, functional materials placed at magnetic anti-nodes or nodes will experience enhanced or zero magnetic excitation, respectively.
If follows from the nature of electromagnetic waves, for which wavevector $k$, electric field $E$ and magnetic field $B$ form a right-handed set, that electric nodes correspond to magnetic anti-nodes and vice versa. This provides a spectroscopic opportunity to excite thin materials either electrically or magnetically and it allows enhancement and complete cancelation of any optical phenomena that are controlled by either electric or magnetic excitation.
Here we apply this concept to several thin optically active metamaterials and an optically anisotropic structure.
% removed references:
% wave plate achiralFishscale
% linearly birefringent/dichroic RPM NComms_2012_THzMalteseCrossRPM
% ... and even reconfigurable planar metamaterials with adjustable properties \cite{NComms_2012_THzMalteseCrossRPM, NNanotech_2013_ElectrostaticRPM}

\begin{figure} [tb]
\includegraphics[width=80mm]{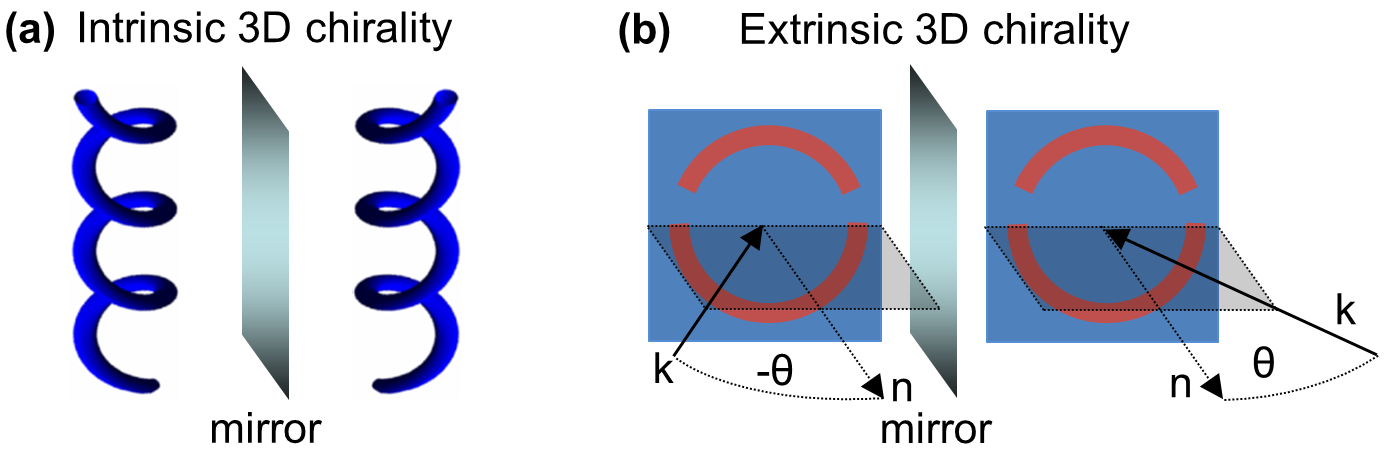}
\caption{\label{Fig_chirality}
\textbf{Intrinsic and extrinsic 3D chirality.} (a) Any object that cannot be superimposed with its mirror image has intrinsic 3D chirality \cite{Kelvin}. (b) Mirror symmetry of the experimental arrangement can also be broken by the light propagation direction. Extrinsic 3D chirality is present when the experimental arrangement consisting of material and direction of incidence cannot be superimposed with its mirror image. Extrinsic 3D chirality occurs at oblique incidence onto planar patterns lacking twofold rotational symmetry if there is no line of mirror symmetry parallel to the plane of incidence (marked by a gray sheet, defined by incidence direction $k$ and surface normal $n$) \cite{APL_Plum_2008_Extr3D, PRL_Plum_2009_Extr3D}. Mirror forms corresponding to the opposite handedness are shown for both cases.
}
\end{figure}

Optical activity manifests itself as circular birefringence and circular dichroism, which rotate the polarization azimuth of electromagnetic waves and change the polarization ellipticity, respectively. Conventionally, optical activity is seen in intrinsically 3D-chiral materials such as twisted organic molecules (e.g. proteins, DNA, sugar molecules), chiral crystals (e.g. quartz, tartaric acid) and artificial helical structures \cite{Pasteur_1848, RSocLon_Bose_1898_RotPlanePolByTwistedStructure, AnnalenDerPhysik_1920_Lindman_helices}, which cannot be superimposed with their mirror image, see Fig.~\ref{Fig_chirality}a. Particularly large optical activity in thin structures has been observed for intrinsically 3D-chiral stereometamaterials based on pairs of identical, mutually twisted metal patterns in parallel planes \cite{APL_Plum_2007_GiantOpticalGyrotropy, PRB_Plum_2009_chiralNIM, PRB_2009_Zhou_TwistedCrosses_EffParameterRetrieval, OL_Decker_2009_TwistedCrosses}. It is less well-known that optical activity can also occur in achiral materials, if the direction of incidence breaks the symmetry of the experimental arrangement \cite{Bunn, Williams}. Such extrinsic 3D chirality, which leads to exceptionally large optical activity in metamaterials \cite{APL_Plum_2008_Extr3D, PRL_Plum_2009_Extr3D}, occurs for oblique incidence onto planar patterns lacking twofold rotational symmetry, when the experimental arrangement consisting of material and direction of incidence becomes different from its mirror image, see Fig.~\ref{Fig_chirality}b. For such patterns extrinsic 3D chirality will only be absent when the structure has a line of mirror symmetry in the plane of incidence.
% Possible theoretical reference APL_Svirko_2001_LayeredChiralMicrostructures

\begin{table}
\caption{Interference of coherent beams at oblique incidence.\label{Table_interference}}
\begin{tabular}{ l | l | l }
                & $E$ anti-node ($\alpha=0^\circ$)           & $E$ node ($\alpha=180^\circ$) \\
  \hline
 p-polarization & $E_{||}=2E_0 \cos(\theta)$    & $E_{||}=0$ \\
                & $B_{||}=0$                   & $B_{||}=2B_0$ \\
                & $E_{\perp}=0$                & $E_{\perp}=2E_0 \sin(\theta)$ \\
                & $B_{\perp}=0$                & $B_{\perp}=0$ \\
  \hline
 s-polarization & $E_{||}=2E_0$                & $E_{||}=0$ \\
                & $B_{||}=0$                   & $B_{||}=2B_0 \cos(\theta)$ \\
                & $E_{\perp}=0$                & $E_{\perp}=0$ \\
                & $B_{\perp}=2B_0 \sin(\theta)$ & $B_{\perp}=0$ \\
\end{tabular}
\end{table}

With respect to extrinsic chirality, it is important to note that
the coherent control concept can also be applied at oblique incidence as the constant phase difference $\alpha$ between both incident beams at the metamaterial plane is preserved for coherent beams incident in the same plane at the same angle $\theta$ on the same side of the metamaterial's surface normal. In this case, the electric and magnetic field components parallel to the metamaterial plane interfere in the same way as at normal incidence ($E_{||}$, $B_{||}$), and new field components normal to the metamaterial plane appear ($E_{\perp}$, $B_{\perp}$) depending on the incident polarization, compare Fig.~\ref{Fig_concept} and Table~\ref{Table_interference}.

\section{Results}

\begin{figure*} [H!]
\includegraphics[width=150mm]{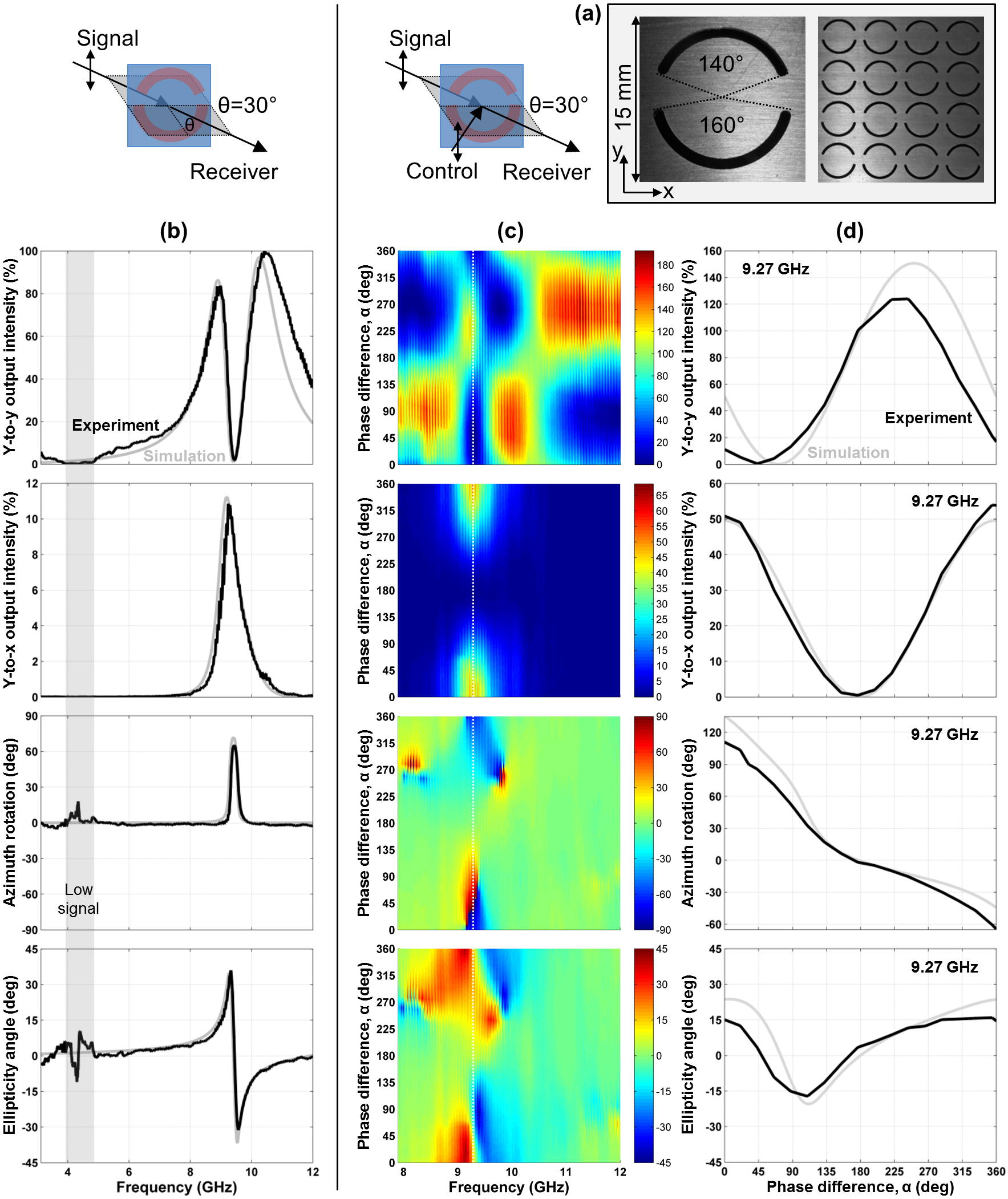}
\caption{\label{Fig_extr_nASR}
\textbf{A coherent polarization rotator based on extrinsic 3D chirality.} (a) Unit cell and fragment of the metamaterial array consisting of asymmetrically split ring apertures in an aluminum sheet. (b) Transmission characteristics of the metamaterial for $30^\circ$ oblique incidence of a y-polarized signal beam in terms of y-to-y intensity transmission, y-to-x intensity conversion, azimuth rotation and ellipticity angle of the detected beam. (c) Coherent control of these optical properties with an additional y-polarized control beam as a function of the phase difference $\alpha$ between the control and signal beams. (d) The same optical properties for a selected frequency of 9.27~GHz, where the metamaterial behaves as a coherent polarization rotator
which can rotate the polarization azimuth continuously over the full $180^\circ$ range while the ellipticity angle remains small (within $\pm15^\circ$). Black and grey solid lines correspond to experimental and simulation data, respectively.
}
\end{figure*}

\begin{figure*} [tb]
\includegraphics[width=150mm]{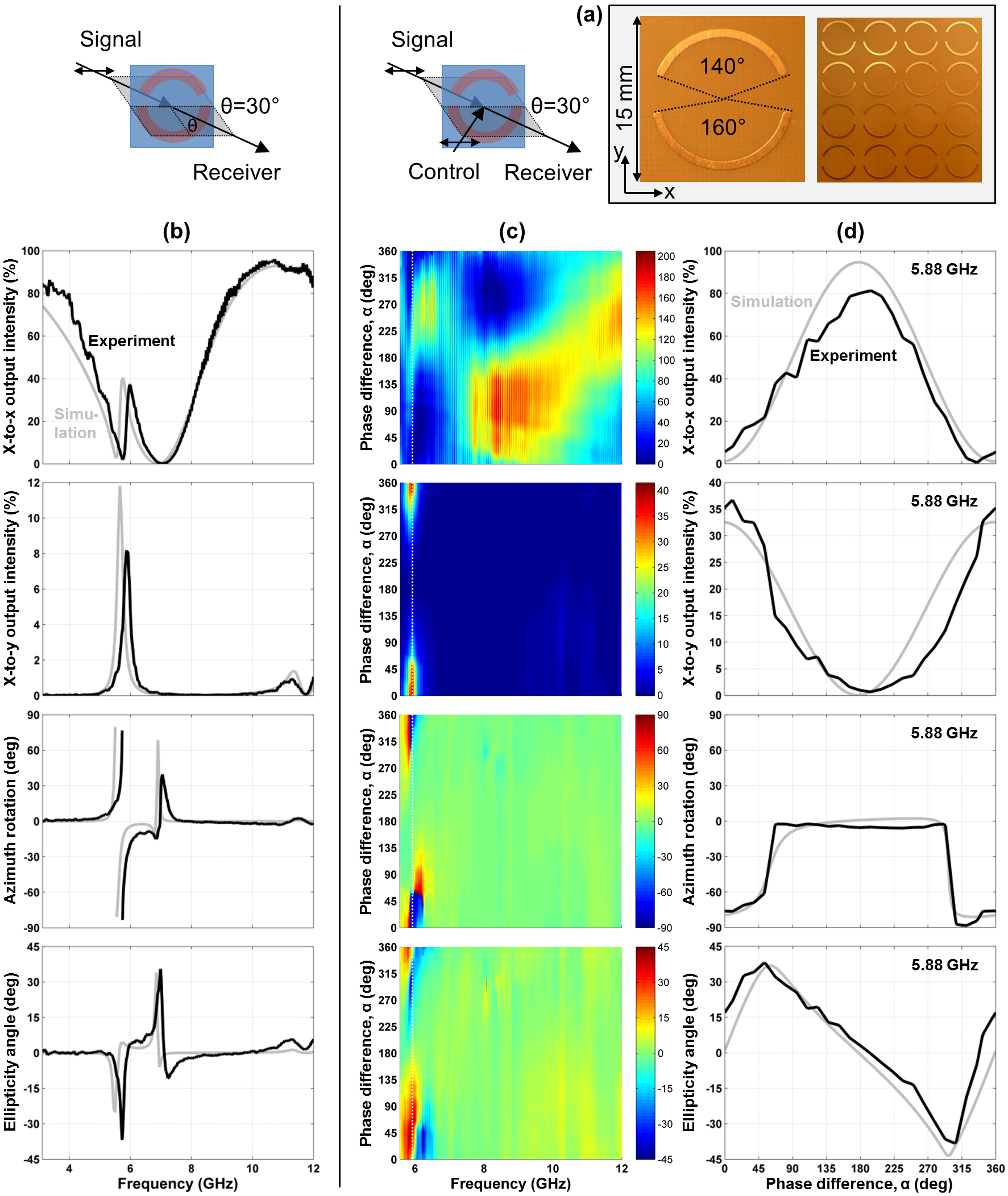}
\caption{\label{Fig_extr_pASR}
\textbf{A coherent ellipticity controller based on extrinsic 3D chirality.}  (a) Unit cell and fragment of the metamaterial array consisting of asymmetrically split wire rings. (b) Transmission characteristics of the metamaterial for $30^\circ$ oblique incidence of an x-polarized signal beam in terms of x-to-x intensity transmission, x-to-y intensity conversion, azimuth rotation and ellipticity angle of the detected beam. (c) Coherent control of these optical properties with an additional x-polarized control beam as a function of the phase difference $\alpha$ between the control and signal beams. (d) The same optical properties for a selected frequency of 5.88~GHz, where the metamaterial behaves as a coherent ellipticity controller. Solid black and gray lines correspond to experimental and simulation data, respectively.
}
\end{figure*}

\begin{figure*} [tb]
\includegraphics[width=150mm]{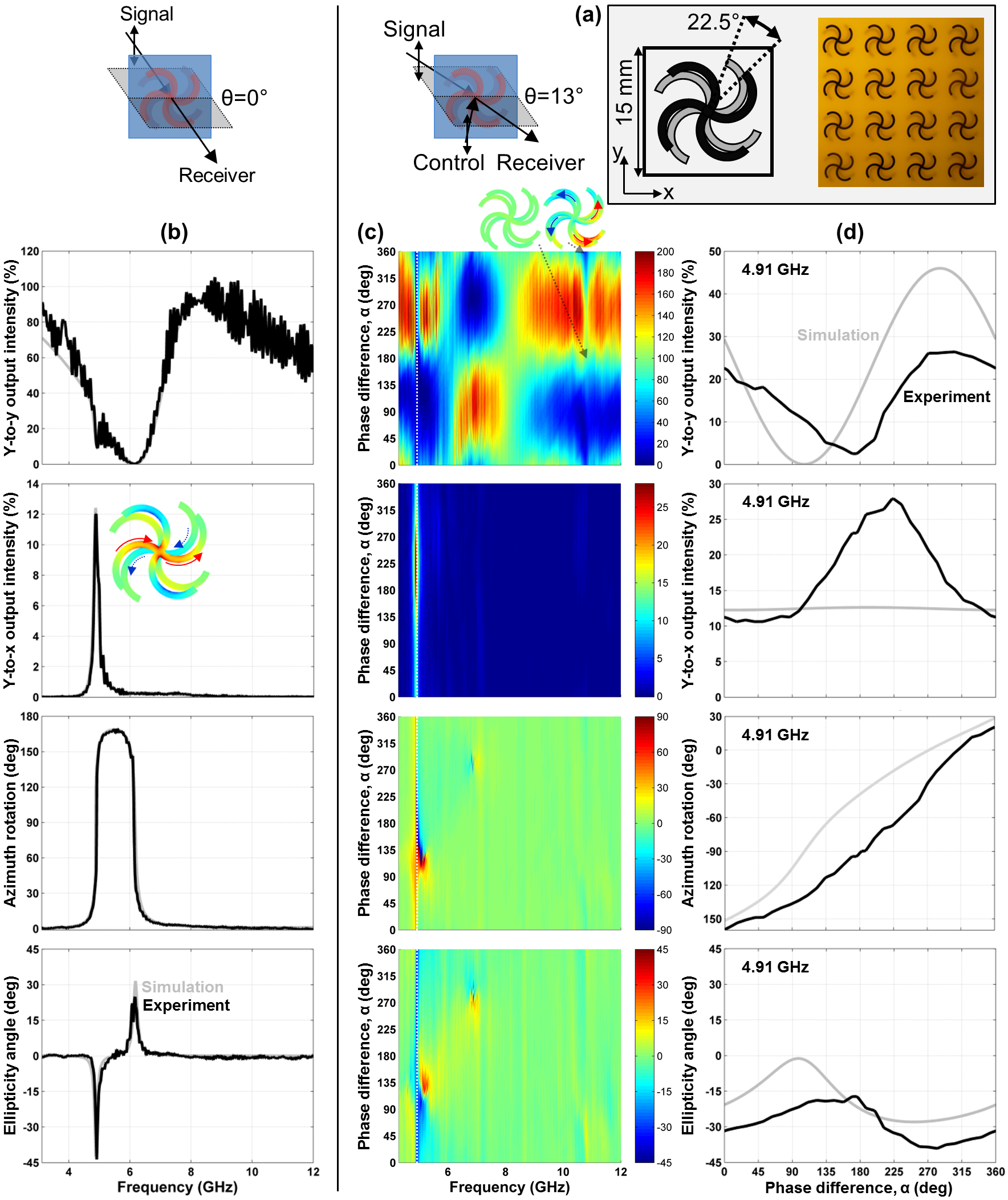}
\caption{\label{Fig_intr}
\textbf{A coherent polarization rotator based on intrinsic 3D chirality.} (a) Unit cell and fragment of the metamaterial array consisting of pairs of mutually twisted metal patterns spaced by a dielectric substrate of 1.6~mm thickness. (b) Transmission characteristics of the metamaterial at normal incidence of a y-polarized signal beam in terms of y-to-y intensity transmission, y-to-x intensity conversion, azimuth rotation and ellipticity angle of the detected beam. The inset shows the $x$-component of the current density at the 4.9~GHz resonance. (c) Coherent control of these optical properties at close to normal incidence ($\theta=13^\circ$) with an additional y-polarized control beam as a function of the phase difference $\alpha$ between the control and signal beams. Insets show the $x$-component of the current density at the 10.7~GHz resonance for phases $\alpha=0^\circ$ and $180^\circ$. (d) The same optical properties for a selected frequency of 4.91~GHz, where the metamaterial behaves as a coherent polarization rotator which can rotate the polarization azimuth continuously over the full $180^\circ$. Solid black and gray lines correspond to experimental and simulation data, respectively.
}
\end{figure*}

We study coherent control of optical activity and optical anisotropy for three metamaterials in the microwave part of the spectrum between 3 and 12~GHz (100-25~mm wavelength). All structures have a diameter of 220~mm, a thickness of no more than 1.6~mm and they consist of a square array of $15\times 15~\text{mm}^2$ meta-molecules, which renders them thin compared to the wavelength and non-diffracting throughout our experiments.
The metamaterials are an array of asymmetrically split ring apertures cut into a 1~mm thick aluminum sheet (Fig.~\ref{Fig_extr_nASR}), the complementary array of asymmetrically split wire rings etched on a 1.6~mm thick dielectric substrate (Fig.~\ref{Fig_extr_pASR}) and an intrinsically 3D-chiral array consisting of mutually twisted pairs of metal patterns etched on both sides of a dielectric substrate of the same thickness (Fig.~\ref{Fig_intr}).

\begin{figure*} [tb]
\includegraphics[width=150mm]{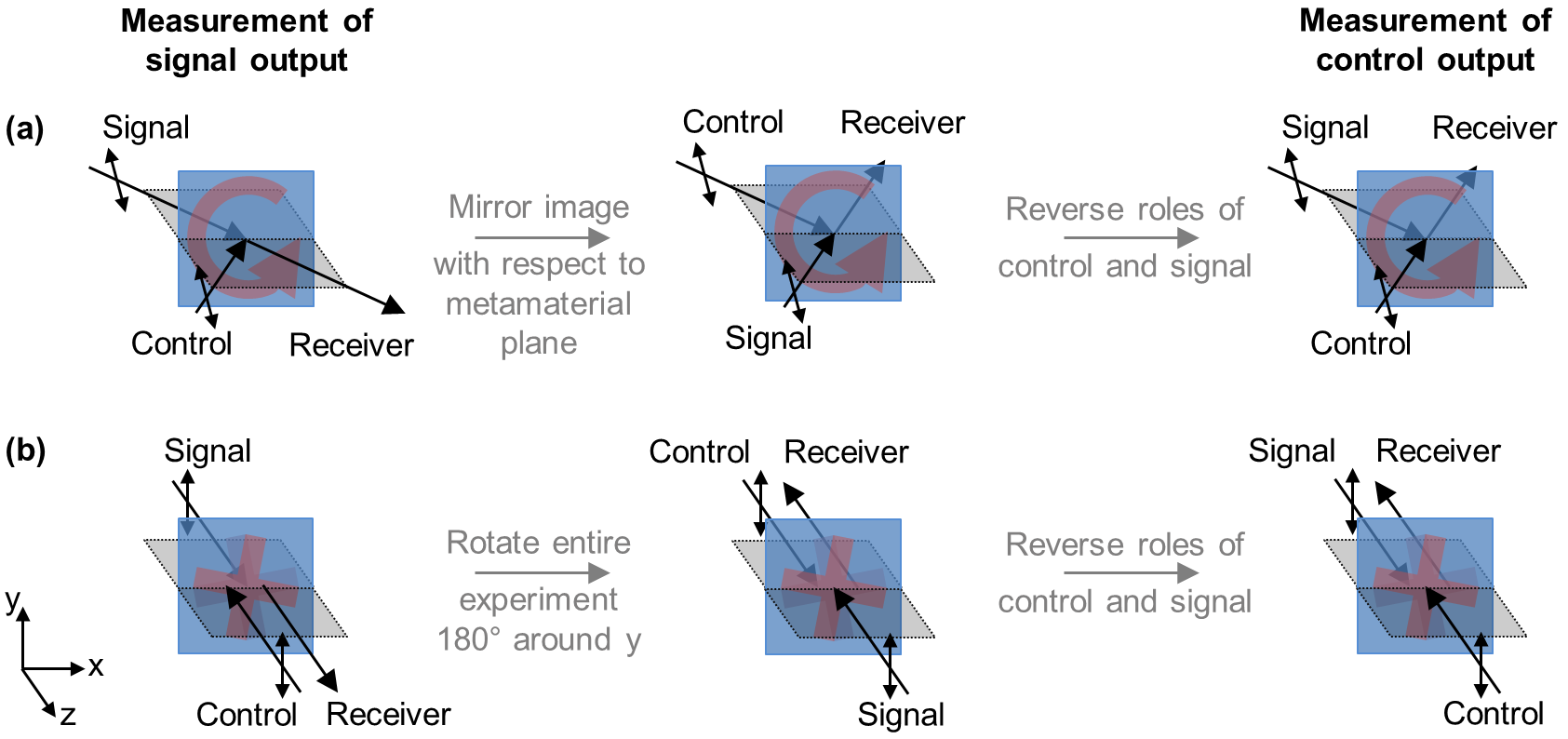}
\caption{\label{Fig_symmetry}
\textbf{Signal output vs control output.} (a) Planar metamaterials: Starting from the signal output for phase $\alpha$, for an arbitrary planar metamaterial and an arbitrary choice of polarization, the control output corresponds to the mirror image experiment (\emph{reversed} azimuth rotation and ellipticity) with reversed roles of signal and control beams (phase $-\alpha$).
(b) The simplest case for intrinsic 3D chirality: Starting from the signal output for phase $\alpha$, for normally incident $y$-polarization onto a pair of mutually twisted crosses in parallel planes, the control output corresponds to the same experiment (\emph{same} azimuth rotation and ellipticity) with reversed roles of signal and control beams (phase $-\alpha$).
This relationship can be generalized to any choice of linear polarization for metamaterials with
polarization azimuth-independent optical properties (e.g. 3-fold or higher rotational symmetry) that are identical for opposite directions of illumination. (a,~b) The metamaterial is represented by a unit cell, the plane of incidence is shown by a gray sheet and the polarization state is indicated by a double arrow. The control beam polarization is chosen so that it has the same projection onto the metamaterial plane as the signal beam polarization.
}
\end{figure*}

\subsection{Coherent control of optical activity due to extrinsic 3D chirality}

The metamaterials based on asymmetrically split ring patterns were studied at oblique incidence where the pattern's line of mirror symmetry was kept perpendicular to the plane of incidence to introduce extrinsic 3D chirality as illustrated by Fig.~\ref{Fig_chirality}b. By choosing an incident polarization parallel or perpendicular to the metamaterial's line of mirror symmetry we avoid polarization effects linked to the linear anisotropy of the split ring pattern.
As shown for the split ring aperture array (s-polarization, Fig.~\ref{Fig_extr_nASR}b) and the split ring wire array (p-polarization, Fig.~\ref{Fig_extr_pASR}b) for an angle of incidence of $\theta=30^\circ$, these metamaterials exhibit large optical activity at their respective resonances. Substantial conversion from the incident linear polarization to the perpendicular polarization of about 10\% of the incident intensity is seen in both cases and the transmitted wave becomes strongly elliptically polarized with azimuth rotation reaching more than $60^\circ$.
No polarization conversion or optical activity was seen at normal incidence and opposite circular birefringence and circular dichroism were observed at opposite angles of incidence $\pm\theta$. The observed optical activity was generally consistent with that reported in refs.~\onlinecite{APL_Plum_2008_Extr3D, PRL_Plum_2009_Extr3D} for similar structures.

Here we study how this optically active metamaterial response is affected by an additional coherent control beam of the same polarization as the incident signal beam, see Figs~\ref{Fig_extr_nASR}c and \ref{Fig_extr_pASR}c. Intensity and polarization of the detected signal output beam strongly depend on the phase difference between the control and signal input beams. For in-phase electrical excitation ($\alpha=0^\circ$) of the metamaterials by the signal and control beams the electric excitation field doubles, leading to a corresponding increase of the structures' scattered fields. This is clearly seen in polarization conversion, which originates from the scattering properties of the optically active metamaterials. In comparison to single beam excitation, the measured y-to-x (or x-to-y) polarization conversion approximately doubles in terms of fields, corresponding to an approximately fourfold increase in terms of intensity. On the other hand, for anti-phase excitation ($\alpha=180^\circ$), no polarization conversion or optical activity has been detected and the signal output intensity is about 100\%. Here, destructive interference of the incident electric fields prevents excitation of the metamaterial structure, effectively rendering it transparent at all frequencies.
The proportionality of metamaterial excitation and scattering to the electric excitation field $E_{||}$, which is controlled by interference of the input beams, is reflected by the sinusoidal phase dependence of the output intensities. In case of polarization conversion this leads to an output intensity proportional to $\cos^2 \tfrac{\alpha}{2}$.
Between the extremes of enhanced and zero coupling to the metamaterials, the relative phase of signal and control beams allows the output azimuth and ellipticity to be controlled over a wide range around the resonances at 9~GHz (Fig.~\ref{Fig_extr_nASR}c) and 6~GHz (Fig.~\ref{Fig_extr_pASR}c).
It is interesting to pick out specific frequencies where the metamaterials exhibit particularly interesting behaviour. At 9.27~GHz, the split ring aperture array acts as a coherent azimuth modulator, which uniquely and approximately linearly maps the phase of the control beam onto the azimuth of the output beam from $-90^\circ$ to $+90^\circ$, while the ellipcity angle of the output beam remains relatively small, see Fig.~\ref{Fig_extr_nASR}d. At 5.88~GHz, the wire split ring metamaterial acts as a coherent ellipticity controller, which controls the ellipticity angle of the output beam continuously from right-handed circular polarization to left-handed circular polarization, see Fig.~\ref{Fig_extr_pASR}d.

While only the signal output beam can be measured in our experiments, it is clear that there is a second output beam in the direction of the transmitted control beam, see Fig.~\ref{Fig_concept}. As illustrated by Fig.~\ref{Fig_symmetry}a, a measurement of the control output beam corresponds to the mirror image experiment with reversed roles of signal and control beams. This implies that the control output at phase $\alpha$ corresponds to the signal output at phase $-\alpha$ or $360^\circ-\alpha$ with reversed signs of polarization azimuth rotation and ellipticity. This symmetry only applies to truly planar metamaterials, which have reflection symmetry with respect to the metamaterial plane, i.e. it should apply to the split ring aperture array, approximately apply to the split ring wire array (where the substrate breaks reflection symmetry with respect to the metamaterial plane) and it will not apply to the intrinsically 3D-chiral metamaterial (which reverses handedness upon reflection on the metamaterial plane).

It follows from these symmetry considerations and our measurements that all polarization effects vanish for both output beams for anti-phase excitation ($\alpha=180^\circ$). In particular, for the split ring aperture array, which is essentially lossless at microwave frequencies, it follows that both output intensities must be 100\% in this case. More generally, for the lossless metamaterial signal and control output intensities (phases $\alpha$ and $360^\circ-\alpha$, respectively) should add up to 200\%, which roughly agrees with our experimental results across the investigated spectral range, see Fig.~\ref{Fig_extr_nASR}c. On the other hand, the wire split ring metamaterial exhibits substantial losses around its resonant region between 5.5 and 7~GHz, where in-phase excitation ($\alpha=0^\circ$) leads to substantial coherent absorption \cite{LSA_2012_MMCoherentAbsorption} in the lossy dielectric substrate, while the total output intensity remains relatively close to 200\% at non-resonant frequencies, see Fig.~\ref{Fig_extr_pASR}c.

Both metamaterials can also act as coherent intensity modulators, which map the relative phase of the signal and control beams onto the intensity of the output beams. For phases $\alpha=90^\circ$ and $\alpha=270^\circ$, the split ring aperture array (Fig.~\ref{Fig_extr_nASR}c, 10~GHz and 10.5-12~GHz) and the wire split ring metamaterial (Fig.~\ref{Fig_extr_nASR}c, 8.4~GHz) direct almost the entire intensity of both input beams into the same output beam without polarization change. Considering a phase modulated signal beam and a reference control beam, this will yield two separate amplified intensity modulated output beams corresponding to the leading and trailing signal phase components, respectively.

\subsection{Coherent control of optical activity due to intrinsic 3D chirality}

The intrinsically 3D-chiral metamaterial based on mutually twisted metal patterns in parallel planes was studied at close to normal incidence ($\theta=0^\circ$ without control beam, $\theta=13^\circ$ with control beam), see Fig.~\ref{Fig_intr}. In contrast to the previous case, the metamaterial's optical properties are polarization azimuth independent at normal incidence due to its fourfold rotational symmetry and only weakly depend on the angle of incidence for small incidence angles.
As shown by Fig.~\ref{Fig_intr}b, the metamaterial exhibits large circular birefringence and circular dichroism near its resonances at about 4.9 and 6~GHz and y-to-x polarization conversion peaks at the 4.9~GHz resonance reaching about 12\% of the incident intensity. The structure's properties are consistent with those described in ref.~\onlinecite{PRB_Plum_2009_chiralNIM}, where a negative index of refraction due to giant optical activity has been reported for a similar metamaterial design.

Here we study how the optically active metamaterial response is affected by an additional coherent control beam of the same polarization as the signal beam, see Fig.~\ref{Fig_intr}c. As the control and receiving antennas had to be placed next to each other, the coherent control experiment had to be conducted at a small angle of incidence of $\theta=13^\circ$, which may be considered a good approximation to normal incidence conditions as explained above. The results shown here correspond to s-polarization, however, the metamaterial's optical properties for p-polarization were found to be almost identical.
Similarly to the previous case, the intensity and polarization state of the signal output beam strongly depend on the phase difference between the input beams. However, in contrast to the planar metamaterials discussed above, there is no special phase $\alpha$ where all polarization effects vanish. In contrast, our measurements and simulations show that y-to-x polarization conversion is non-zero at all phases and it peaks in our experiments near the magnetic anti-node, where electric excitation of the metamaterial vanishes ($\alpha=180^\circ$). This suggests that magnetic coupling to the metamaterial plays an important role, and indeed, the 4.9~GHz resonance has previously been identified as a magnetic resonance leading to a negative effective permeability \cite{PRB_Plum_2009_chiralNIM} and is characterized by counterpropagating currents in the metal patterns on opposite sides of the substrate (inset to Fig.~\ref{Fig_intr}b). Such a magnetic resonance can be excited effectively by the magnetic anti-node, where the magnetic field oscillates with maximum amplitude within the metamaterial plane. The metamaterial's response is particularly interesting at 4.9~GHz, where the structure acts as a coherent polarization rotator that uniquely maps the phase $\alpha$ onto the polarization azimuth of the (strongly elliptically polarized) output beam, see Fig.~\ref{Fig_intr}d.

In contrast to planar metamaterials, the signal output and control output beams are more simply related in case of intrinsically 3D-chiral metamaterials with polarization azimuth-independent optical properties that are identical for opposite directions of illumination.
As the 3D-chiral twist is the same for observation from opposite sides, at normal incidence, a measurement of the control output is equivalent to simply interchanging the roles of signal and control beams, as illustrated for the simplest case of pairs of mutually twisted crosses \cite{PRB_2009_Zhou_TwistedCrosses_EffParameterRetrieval, OL_Decker_2009_TwistedCrosses} in Fig.~\ref{Fig_symmetry}b.
This implies that the control output at phase $\alpha$ corresponds to the signal output at phase $-\alpha$ or $360^\circ-\alpha$ with the same signs of polarization azimuth rotation and ellipticity.

Our experimental 2D-chiral crosses of fourfold rotational symmetry exhibit polarization azimuth-independent optical properties and complete absence of 2D-chiral optical effects at normal incidence \cite{PRB_Plum_2009_chiralNIM, PRL_Fedotov_2006_AsymmetricTransmissionMW}. However, slightly oblique incidence and slightly different resonant reflectivity for opposite directions of illumination resulting from the different orientation of the metal patterns on opposite sides of the dielectric substrate make the above symmetry an approximation near resonances in our experimental case.

Considering the relationship between signal and control output beams, it is clear from Fig.~\ref{Fig_intr}c that absorption is generally low in the non-resonant regions, where the output intensities of the signal ($\alpha$) and control ($360^\circ-\alpha$) output beams add up to about the total input intensity of 200\%, with roughly 100\% intensity in each output beam for $\alpha=180^\circ$, very much in the same way as for the planar metamaterials discussed above. However, the situation is very different at the 4.9~GHz resonance, where absorption is large for all values of $\alpha$, indicating that both electric and magnetic excitation fields, $E_{||}$ and $B_{||}$, strongly couple to the metamaterial.
Similarly to the previous cases, bands of coherent intensity modulation without polarization change can be seen in non-resonant regions around 7~GHz and 10~GHz, where the metamaterial directs almost the entire intensity of both input beams into a single output beam for phases $\alpha=90^\circ$ and $270^\circ$.

\begin{figure*} [tb]
\includegraphics[width=150mm]{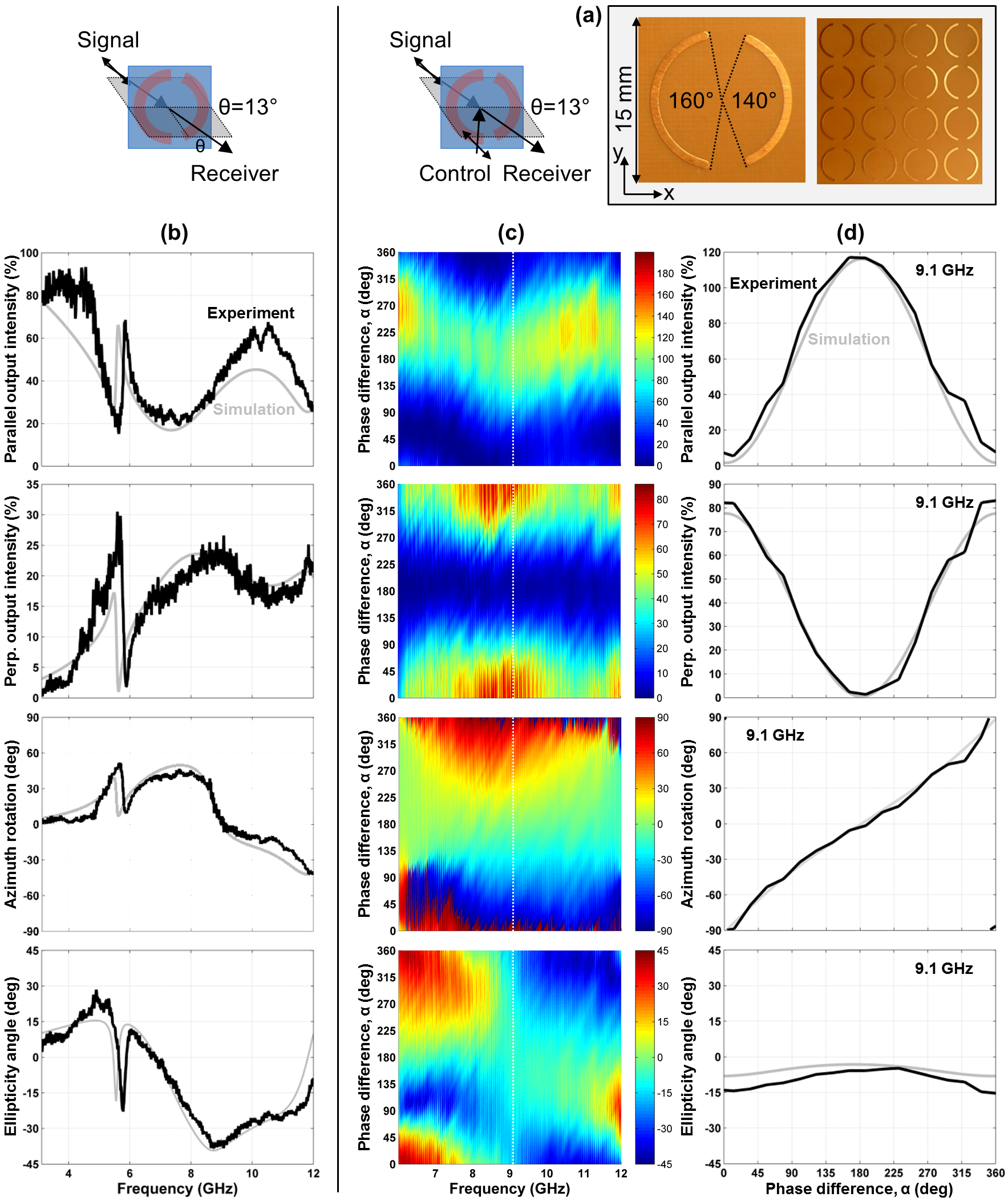}
\caption{\label{Fig_anisotropy}
\textbf{A coherent polarization rotator based on anisotropy.} (a) Unit cell and fragment of the metamaterial array consisting of asymmetrically split wire rings. (b) Transmission characteristics of the metamaterial for $13^\circ$ oblique incidence of a diagonally polarized signal beam in terms of transmitted intensities polarized parallel and perpendicular to the incident polarization, azimuth rotation and ellipticity angle. (c) Coherent control of these optical properties with an additional control beam polarized parallel to the signal beam as a function of the phase difference $\alpha$ between the control and signal beams. (d) The same optical properties for a selected frequency of 9.1~GHz, where the metamaterial behaves as a coherent polarization rotator which can rotate the polarization azimuth continuously over the full $180^\circ$ range while the ellipticity angle remains small (within $\pm15^\circ$). Solid black and gray lines correspond to experimental and simulation data, respectively.
}
\end{figure*}

\subsection{Coherent control of optical anisotropy}

Optical anisotropy manifests itself as linear birefringence and linear dichroism, leading to phase delays and differential transmission of the ordinary and extraordinary linear eigenpolarizations of optically anisotropic materials, respectively. Optical anisotropy is found in many crystals (e.g. calcite, quartz and magnesium fluoride) and linear birefringence is exploited in waveplates, while linear dichroism is the basis of thin film linear polarizers.

While suitable directions of incidence onto the split ring metamaterials lead to extrinsic chirality, as discussed above, these structures also exhibit linear optical anisotropy. This is particularly apparent for achiral experimental arrangements such as normal incidence, where the ordinary and extraordinary polarizations are oriented parallel and perpendicular to the pattern's line of mirror symmetry. As for linearly birefringent crystals, the ordinary and extraordinary polarizations will not be changed by the metamaterial, but the polarization effects associated with optical anisotropy can be easily studied for illumination with an intermediate polarization state. Therefore we study optical anisotropy for the wire split ring metamaterial by illuminating it with waves polarized at $45^\circ$ to its line of mirror symmetry under quasi-normal incidence conditions (an achiral configuration where the pattern's line of mirror symmetry coincides with the plane of incidence with $\theta=13^\circ$). This is just like one would illuminate a linearly birefringent quarter waveplate to create circular polarization.
As illustrated by Fig.~\ref{Fig_anisotropy}b, under these illumination conditions, the anisotropic metamaterial dramatically changes the azimuth and ellipticity angle of the transmitted beam across almost the entire investigated spectral range. In particular, it behaves like a quarter waveplate around 9~GHz, where the transmitted beam has left-handed circular polarization.

Here we study how the anisotropic metamaterial response is affected by an additional coherent control beam of the same polarization as the signal beam, see Fig.~\ref{Fig_anisotropy}c. Similarly to the extrinsically chiral configurations discussed above, both polarization components of the signal output intensity have a sinusoidal phase dependence and the metamaterial is essentially transparent with 100\% signal output and complete absence of polarization changes at the electric node ($\alpha=180^\circ$). Also, polarization conversion in terms of intensity is proportional to $\cos^2 \tfrac{\alpha}{2}$, peaking for in-phase electrical excitation ($\alpha=0^\circ$) with approximately fourfold increased values compared to absence of the control beam.
However, in contrast to the examples of resonant optical activity, the anisotropic polarization conversion and the associated polarization azimuth rotation and ellipticity changes are non-resonant and therefore broadband and low loss, despite the presence of the lossy dielectric substrate.
As a result, the metamaterial acts as a broadband coherent polarization rotator that uniquely maps phase onto polarization azimuth from 6.5~GHz to 11.5~GHz. This is illustrated in detail by Fig.~\ref{Fig_anisotropy}d for 9.1~GHz where the ellipticity of the output polarization remains small. Similarly, the structure acts as a coherent ellipticity modulator between 6 and 7~GHz, which allows the output beam to be continuously tuned from right-handed circular polarization to left-handed almost circular polarization.
As explained above based on Fig.~\ref{Fig_symmetry}a for planar metamaterials in general, the control output beam at phase $\alpha$ corresponds to the signal output beam at phase $360^\circ-\alpha$ with reversed signs of polarization azimuth rotation and ellipticity. However, the wire split ring array is not perfectly planar due to the presence of the substrate, which leads to slight deviations from this relationship between the signal and control output beams. In particular, simulations confirm that the signal output of slightly above 100\% for $\alpha=180^\circ$ is accounted for by a control output of slightly below 100\% intensity as one would expect from energy conservation.

\section{Discussion}

For the planar metamaterials, transparency and complete absence of all polarization effects at the magnetic anti-node ($\alpha=180^\circ$) in all experimental configurations (for both s and p polarizations at normal and oblique incidence for both samples) indicates that tangential magnetic $B_{||}$ and normal electric $E_{\perp}$ fields cannot excite the planar structures, which should be expected where the motion of electrical charges is confined to a single plane (see Table~\ref{Table_interference}). On the other hand, all planar structures were strongly excited by tangential electric fields $E_{||}$ leading to large resonant optical activity in extrinsically 3D-chiral experimental configurations (where the excitation field was orientated parallel or perpendicular to the pattern's line of symmetry to avoid optical anisotropy effects) and broadband polarization effects resulting from optical anisotropy for achiral quasi-normal incidence illumination with a polarization state containing both ordinary and extraordinary polarization components.

In contrast to the planar structures, resonant optical activity at quasi-normal incidence was observed for a layered intrinsically 3D-chiral structure both at the electric and magnetic anti-nodes, indicating that the optically active resonance at 4.9~GHz can be excited by both tangential electric $E_{||}$ and magnetic $B_{||}$ fields.
Measurements for s and p polarizations at small angles of incidence did yield almost identical results, suggesting that the associated smaller magnetic and electrical excitation normal to the metamaterial plane does not strongly affect the metamaterial response.
(Except for the absorption resonance without significant polarization changes at 10.7~GHz for $\alpha=0^\circ$ [see insets of Fig.~\ref{Fig_intr}c], which is only seen for s-polarization, indicating that it originates from $B_\perp$ excitation [compare Table~\ref{Table_interference}].)
% seen for s-polarization in coherent control experiment, oblique incidence single beam experiment and single beam simulation

The different relationship between signal output and control output for the intrinsically 3D-chiral metamaterial and the planar metamaterials can be easily understood as follows (compare Fig.~\ref{Fig_symmetry}). For all structures, a measurement of the control output resembles a measurement of the signal output for interchanged input beams, i.e. for the opposite phase difference between signal and control beams.
However, while intrinsic 3D chirality is the same for both input beams, for extrinsically 3D-chiral configurations the handedness associated with the directions of signal and control beams is opposite. Similarly, in case of optical anisotropy, where polarization effects occur for incident polarizations that are neither parallel nor perpendicular to the planar pattern's line of mirror symmetry, an excitation polarization azimuth to the right of the pattern's mirror line as seen looking into the signal beam, corresponds to an azimuth to the left of the mirror line when looking into the control beam. Therefore, the control output corresponds to the signal output for the opposite phase with \emph{same} azimuth rotation and ellipticity for intrinsic 3D chirality and \emph{opposite} azimuth rotation and ellipticity for planar metamaterials (extrinsic 3D chirality and anisotropy). These relationships have been confirmed by numerical simulations.

While the experimental results reported here were measured in the microwave part of the spectrum, the same concepts can be easily applied across the electromagnetic spectrum. Optical activity due to intrinsic and extrinsic 3D chirality has also been observed in THz \cite{OE_Singh_2010_THzExtr3D} and optical metamaterials \cite{PRL_KuwataGonokami_2005_GiantOpticalActivity, APL_Plum_2007_GiantOpticalGyrotropy, OL_Decker_2007_CircDichroism, PRL_Plum_2009_Extr3D, OL_Decker_2009_TwistedCrosses} of sub-wavelength thickness and numerous thin anisotropic metamaterial patterns have been reported in the literature \cite{WireGrid_visible, OE_2009_THz_QuarterWavePlates}, including a photonic metamaterial for which coherent control of absorption has been reported recently \cite{LSA_2012_MMCoherentAbsorption}. Coherent excitation of these structures in their respective spectral ranges of operation should lead to similar optical phenomena to those reported here.

Coherent control of polarization states and intensity has tremendous potential for ultrafast modulation of electromagnetic waves.
For non-resonant polarization control (as reported here for optical anisotropy) and non-resonant intensity modulation (as seen in most of our experiments), achievable modulation rates should approach the frequency of the electromagnetic wave itself, while achievable modulation rates for resonant polarization control (as reported here for optical activity) should be lower by about the resonance quality factor. Applied to the optical telecommunications band around $1.5~\mu$m, where metamaterial resonances typically have quality factors on the order of 10, this promises modulation rates on the order of 10~THz for resonant effects.
% 1.5um = 200 THz

More generally, by allowing the selective material excitation with tangential electric or magnetic fields, coherent control provides a useful tool for probing thin materials at normal or oblique incidence. Also selective probing with electric and magnetic fields perpendicular to the interface is possible with incident p and s-polarizations, however, it comes at the cost that only one field component vanishes parallel to the interface, compare Table~\ref{Table_interference}.

\section{Conclusions}

In summary, we show that coherent control of functional materials of subwavelength thickness allows ultrafast manipulation of their polarization properties.  Coherent control may be achieved by placing the functional material into the standing wave of two counterpropagating coherent beams, where the excitation in the plane of the functional material can be shifted continuously from electric in the anti-node of the electric field to magnetic in the magnetic anti-node by controlling the relative phase of the illuminating beams.

Specifically, we demonstrate for the first time coherent control of optical activity and optical anisotropy.
We report coherent control of optical activity of an intrinsically 3D-chiral metamaterial, optical activity of extrinsically 3D-chiral planar metamaterials, and optical anisotropy of a planar metamaterial in the microwave part of the spectrum, demonstrating several coherent polarization rotators and coherent ellipticity controllers. We also demonstrate coherent intensity modulators that can direct almost all input intensity into a single output beam. All observed phenomena can be easily scaled throughout the electromagnetic spectrum up to the optical spectral range.
In particular, non-resonant coherent polarization modulators based on optical anisotropy are promising for practical applications due to their low absorption and large bandwidth. Scaled to optical frequencies, such structures promise energy-efficient, ultrafast polarization control with 10s of THz bandwidth for coherent optical data processing networks.

Furthermore, coherent illumination provides means of selectively exciting thin materials with tangential or perpendicular components of electric or magnetic field, opening up a new opportunity for spectroscopy.

\section{Methods}

\subsection{Experimental characterization.}
All samples were studied in a microwave anechoic chamber using 3 broadband microwave antennas and a vector network analyzer (Agilent E8364B). In coherent control experiments, both control and signal antennas (Schwarzbeck BBHA 9120D) were equipped with collimating lenses and were fed with the same signal through identical cables. The minimum angle of incidence of $\theta=13^\circ$ resulted from the physical size of control antenna (Schwarzbeck BBHA 9120D) and receiving antenna (Schwarzbeck STLP 9148), which had to be placed next to each other.

\subsection{Numerical modeling.}
Simulations were conducted using a full three-dimensional Maxwell finite element method
solver in the frequency domain. In case of extrinsic 3D chirality, the angle of incidence in the simulations was $\theta=30^\circ$, just like in the experiments (Figs~\ref{Fig_extr_nASR}b,d and \ref{Fig_extr_pASR}b,d).
In case of the (quasi-)normal incidence experiments for intrinsic 3D chirality and optical anisotropy, the simulations correspond to $\theta=0^\circ$ (Figs~\ref{Fig_intr}b,d and \ref{Fig_anisotropy}b,d).
Modeled and experimental metamaterial resonance frequencies are close, but slightly shifted relative to each other.
Therefore we show the simulated phase dependence of the signal output beam intensity and polarization for the spectral feature that corresponds to the experimental data, which is within 4\% of the experimental frequency in all cases (Figs~\ref{Fig_extr_nASR}d, \ref{Fig_extr_pASR}d, \ref{Fig_intr}d, \ref{Fig_anisotropy}d).

\subsection{Definitions.}
The phase difference $\alpha$ between control an signal beams is defined as increasing with increasing optical length of the control beam path, where in-phase excitation corresponds to the anti-node of electric field in the metamaterial plane. Counterclockwise polarization azimuth rotation as seen looking into the beam is defined as positive. A polarization state is defined as right-handed with positive ellipticity angle when the electric field vector at a fixed point in space rotates clockwise as seen by an observer looking into the beam. The input intensities of the signal and control beams are considered 100\%, each, so that the total output intensity is 200\% in absence of absorption.

\section{Acknowledgements}

The authors are grateful to Xu Fang and Ming Lun Tseng for fruitful discussions. This
work is supported by the MOE Singapore (grant MOE2011-T3-1-005), the Leverhulme Trust, the
Royal Society and the UK's Engineering and Physical Sciences
Research Council through the Nanostructured Photonic Metamaterials Programme Grant.

%\section{Author contributions}

%N.I.Z.~and E.P.~conceived the idea of the experiment, wrote the paper and supervised the work;
%E.P.~designed the experiment; S.A.M~carried out the measurements;
%J.S.~performed numerical modelling;
%all authors discussed the results and analysed the data.

%\textbf{Competing financial interests:} The authors declare no
%competing financial interests.

\bibliographystyle{nature}
\bibliography{coherent_opt_act}

\begin{thebibliography}{10}

\bibitem{LSA_2012_MMCoherentAbsorption}
Zhang, J., MacDonald, K.~F., and Zheludev, N.~I.
\newblock {\em Light: Science and Applications}{ \bf 1}, e18 (2012).

\bibitem{WireGrid_visible}
Ahn, S.~W., Lee, K.~D., Kim, J.~S., Kim, S.~H., JD, J. D.~P., Lee, S.~H., and
  Yoon, P.~W.
\newblock {\em Nanotechnology}{ \bf 16}(9), 1874 (2005).

\bibitem{OE_2009_THz_QuarterWavePlates}
Strikwerda, A.~C., Fan, K., Tao, H., Pilon, D.~V., Zhang, X., and Averitt,
  R.~D.
\newblock {\em Opt. Express}{ \bf 17}, 136--149 (2009).

\bibitem{PRL_KuwataGonokami_2005_GiantOpticalActivity}
Kuwata-Gonokami, M., Saito, N., Ino, Y., Kauranen, M., Jefimovs, K., Vallius,
  T., Turunen, J., and Svirko, Y.
\newblock {\em Phys. Rev. Lett.}{ \bf 95}, 227401 (2005).

\bibitem{APL_Plum_2007_GiantOpticalGyrotropy}
Plum, E., Fedotov, V.~A., Schwanecke, A.~S., Zheludev, N.~I., and Chen, Y.
\newblock {\em Appl. Phys. Lett.}{ \bf 90}, 223113 (2007).

\bibitem{OL_Decker_2007_CircDichroism}
Decker, M., Klein, M.~W., Wegener, M., and Linden, S.
\newblock {\em Opt. Lett.}{ \bf 32}, 856 (2007).

\bibitem{APL_Plum_2008_Extr3D}
Plum, E., Fedotov, V.~A., and Zheludev, N.~I.
\newblock {\em Appl. Phys. Lett.}{ \bf 93}, 191911 (2008).

\bibitem{PRL_Plum_2009_Extr3D}
Plum, E., Liu, X.-X., Fedotov, V.~A., Chen, Y., Tsai, D.~P., and Zheludev,
  N.~I.
\newblock {\em Phys. Rev. Lett.}{ \bf 102}, 113902 (2009).

\bibitem{PRB_Plum_2009_chiralNIM}
Plum, E., Zhou, J., Dong, J., Fedotov, V.~A., Koschny, T., Soukoulis, C.~M.,
  and Zheludev, N.~I.
\newblock {\em Phys. Rev. B}{ \bf 79}, 035407 (2009).

\bibitem{PRB_2009_Zhou_TwistedCrosses_EffParameterRetrieval}
Zhou, J., Dong, J., Wang, B., Koschny, T., Kafesaki, M., and Soukoulis, C.~M.
\newblock {\em Phys. Rev. B}{ \bf 79}, 121104(R) (2009).

\bibitem{OL_Decker_2009_TwistedCrosses}
Decker, M., Ruther, M., Kriegler, C.~E., Zhou, J., Soukoulis, C.~M., Linden,
  S., and Wegener, M.
\newblock {\em Opt. Lett.}{ \bf 34}, 2501 (2009).

\bibitem{Science_2011_Capasso_PhaseGradients}
Yu, N., Genevet, P., Kats, M.~A., Aieta, F., Tetienne, J.-P., Capasso, F., and
  Gaburro, Z.
\newblock {\em Science}{ \bf 334}, 333--337 (2011).

\bibitem{LSA_2013_visible_phase_gradient_lens}
Ni, X., Ishii, S., Kildishev, A.~V., and Shalaev, V.~M.
\newblock {\em Light: Science and Applications}{ \bf 2}, e72 (2013).

\bibitem{Kelvin}
Kelvin, L.
\newblock {\em Baltimore Lectures on Molecular Dynamics and the Wave Theory of
  Light},  619.
\newblock C.J. Clay and Sons, Cambridge University Press Warehouse, London
  (1904).

\bibitem{Pasteur_1848}
Pasteur, L.
\newblock {\em C. R. Acad. Sci. Paris}{ \bf 26}, 535 (1848).

\bibitem{RSocLon_Bose_1898_RotPlanePolByTwistedStructure}
Bose, J.~C.
\newblock {\em Proceedings of the Royal Society of London}{ \bf 63}, 146--152
  (1898).

\bibitem{AnnalenDerPhysik_1920_Lindman_helices}
Lindman, K.~F.
\newblock {\em Ann. Phys.}{ \bf 368}(23), 621 (1920).

\bibitem{Bunn}
Bunn, C.~W.
\newblock {\em Chemical Crystallography}, ~88.
\newblock Oxford University Press, New York (1945).

\bibitem{Williams}
Williams, R.
\newblock {\em Phys. Rev. Lett.}{ \bf 21}, 342--344 (1968).

\bibitem{PRL_Fedotov_2006_AsymmetricTransmissionMW}
Fedotov, V.~A., Mladyonov, P.~L., Prosvirnin, S.~L., Rogacheva, A., Chen, Y.,
  and Zheludev, N.~I.
\newblock {\em Phys. Rev. Lett.}{ \bf 97}, 167401 (2006).

\bibitem{OE_Singh_2010_THzExtr3D}
Singh, R., Plum, E., Zhang, W., and Zheludev, N.~I.
\newblock {\em Opt. Express}{ \bf 18}, 13425 (2010).

\end{thebibliography}

\end{document}